# Coupled-Double-Quantum-Dot Environmental Information Engines: A Numerical Analysis

Katsuaki Tanabe*

*Department of Chemical Engineering, Kyoto University, KU-Katsura, Nishikyo, Kyoto 615-8510, Japan*

*Email: tanabe@cheme.kyoto-u.ac.jp

We conduct numerical simulations for an autonomous information engine comprising a set of coupled double quantum dots using a simple model. The steady-state entropy production rate in each component, heat and electron transfer rates are calculated via the probability distribution of the four electronic states from the master transition-rate equations. We define an information-engine efficiency based on the entropy change of the reservoir, implicating power generators that employ the environmental order as a new energy resource. We acquire device-design principles, toward the realization of corresponding practical energy converters, including that (1) higher energy levels of the detector-side reservoir than those of the detector dot provide significantly higher work production rates by faster states' circulation, (2) the efficiency is strongly dependent on the relative temperatures of the detector and system sides and becomes high in a particular Coulomb-interaction strength region between the quantum dots, and (3) the efficiency depends little on the system dot's energy level relative to its reservoir but largely on the antisymmetric relative amplitudes of the electronic tunneling rates.

## 1. Introduction

In recent years, there has been a remarkable progress in the field of information thermodynamics[1-8] in relation to stochastic thermodynamics out of the equilibrium.[9-14] The exploration of information engines is given a particular interest in this field.[15-18] Bipartite, four-state configurations are a handy model to investigate for a further understanding of information thermodynamics.[19-26] Quantum dots, often referred to as artificial atoms, are an adopted candidate for a material component of such setups. The discrete density of states in quantum dots enables high-performance optoelectronic devices[27-29] as well as single-electron manipulation,[30-32] which makes the discussions clear in information thermodynamics. Therefore, quantum-dot-based information engines as well as refrigerators have been theoretically[20,21,23,25,26,32] and experimentally[33-36] investigated.

To study the operation and performance of engines, the efficiency is one of the most crucial parameters to investigate.[26,37-43] In the present study, we have conducted numerical simulations of an autonomous information engine comprising a set of coupled double quantum dots, by adopting a relatively simple model proposed by Horowitz and Esposito,[23] particularly focusing on the efficiencies of the engines to obtain design principles for the realization of practical devices of this type.

## 2. Theory and Calculation Methods

The model setup of an autonomous information engine studied in this article is based on Ref. 23 and comprises two quantum dots and three thermal/electronic reservoirs around as schematically depicted in Fig. 1. Each quantum dot can contain up to one electron. One quantum dot with an electronic potential energy $\varepsilon_X$ functions as an electronic detector by checking whether an electron is in the other quantum dot through capacitive interaction strength or Coulomb interaction energy $U$ between the two quantum dots. This "detector dot" is kept at a temperature $T_D$ and connected to thermal and electronic reservoirs, both having the same temperature $T_D$ and an electronic potential energy $\mu_D$. The other quantum dot with an electronic potential energy $\varepsilon_Y$ is connected to two reservoirs through electrical leads and enables an electrical current flow. This "system dot" is kept at a temperature $T_S$ and connected to two thermal and electronic reservoirs both at $T_S$ with electronic potential energies $\mu_H$ and $\mu_L$ ($\mu_H > \mu_L$). The potential-energy relations among the components in the setup are schematically shown in Fig. 2 for clarification. By properly setting the transition or tunneling rates across the interfaces between the quantum dots and reservoirs, this double-quantum-dot configuration as a whole can drive electrical current in the direction from the reservoir of $\mu_L$ to that of $\mu_H$ through the system dot against the potential slope and thus generate work as will be shown by our following calculations. Each quantum dot has an electronic state 0 or 1, where 1 and 0 mean that the dot is filled or not filled (i.e., empty) with an electron, respectively. In this way, the total electronic state $(x, y)$ will be $(0, 0)$, $(0, 1)$, $(1, 0)$, or $(1, 1)$, as schematically drawn in Fig. 3. For the state $(1, 1)$, the electronic potential energy in the quantum dots will increase to $\varepsilon_X + U$ and $\varepsilon_Y + U$ for the detector and system dots, respectively, due to Coulomb repulsion. We set the time resolution fine enough so that no simultaneous or diagonal jump, such as a transition from $(0, 0)$ to $(1, 1)$, is assumed in our bipartite setup.

The time evolution of the probability of the states $p(x, y)$ can be generally written as a master equation:

$$d_t p(x, y) = \sum_{x',y'} \{W_{x,x'}^{y,y'} p(x', y') - W_{x',x}^{y',y} p(x, y)\}, \quad (1)$$

where $W_{x,x'}^{y,y'}$ is the transition rate from a state $(x', y')$ to $(x, y)$ and we have:

$$W_{10}^{y} = \Gamma f_y, \quad (2)$$

and

$$W_{01}^{y} = \Gamma(1 - f_y) \quad (3)$$

for the electron transfer on the detector dot and

$$W_x^{10,\upsilon} = \Gamma_x^{\upsilon} f_x^{\upsilon}, \quad (4)$$

and

$$W_x^{01,\upsilon} = \Gamma_x^{\upsilon}(1 - f_x^{\upsilon}) \quad (5)$$

for the system dot in this model. Note again that for the jumps, either $x$ or $y$ is fixed at each time step. $\Gamma$ is the electronic tunneling rate between the detector dot and its reservoir. We assume the density of states in the detector-side reservoir to be uniform so that $\Gamma$ is independent of $y$. $\Gamma_x^{\upsilon}$ is the tunneling rate between the system dot and its reservoirs where $\upsilon$

= $H$ or $L$ corresponds to the higher- or lower-potential reservoir, respectively. In contrast, we assume nonuniform profiles of the density of states in the system-side reservoirs so that $\Gamma_x^\upsilon$ depends on $x$. Fermi distribution functions for the detector and system dots have forms of:

$$f_y = \frac{1}{1+\exp\left(\frac{\varepsilon_X + yU - \mu_D}{T_D}\right)} \quad (6)$$

and

$$f_x^\upsilon = \frac{1}{1+\exp\left(\frac{\varepsilon_Y + xU - \mu_\upsilon}{T_S}\right)}, \quad (7)$$

respectively. For simplicity, the Boltzmann constant is set to unity or absorbed into the temperatures throughout this paper. We then determine the steady-state probability distribution of $p(x, y)$.

The heat flow rate in the direction from the system dot to the detector dot or the counter-clockwise probability circulating rate in the central square edged by the four electronic states in Fig. 3 is then:

$$\begin{aligned} J_h &= W_{10}^0 p(0,0) - W_{01}^0 p(1,0) \\ &= \left(W_1^{10,H} + W_1^{10,L}\right)p(1,0) - \left(W_1^{01,H} + W_1^{01,L}\right)p(1,1) \\ &= W_{01}^1 p(1,1) - W_{10}^1 p(0,1) \\ &= \left(W_0^{01,H} + W_0^{01,L}\right)p(0,1) - \left(W_0^{10,H} + W_0^{10,L}\right)p(0,0). \quad (8) \end{aligned}$$

(n.b., $J_h U$ corresponds to the heat flow in energy per time for a more common definition.) The electric current in the direction from the lower- to higher-potential reservoir of the system dot through the system dot, or the sum of the two counter-clockwise circulating transfer rates in the left and right circles in Fig. 3, is given by:

$$J_e = W_0^{10,L} p(0,0) - W_0^{01,L} p(0,1) + W_1^{10,L} p(1,0) - W_1^{01,L} p(1,1)$$
$$= W_0^{01,H} p(0,1) - W_0^{10,H} p(0,0) + W_1^{01,H} p(1,1) - W_1^{10,H} p(1,0). \quad (9)$$

The total entropy production rate of the whole setup is given by:

$$\dot{S}_{TOT} = -J_e \frac{\Delta\mu}{T_S} + J_h \left( \frac{U}{T_D} - \frac{U}{T_S} \right), \quad (10)$$

where $\Delta\mu \equiv \mu_H - \mu_L$. The total entropy production rate $\dot{S}_{TOT}$ can be split into the entropy production rate in the detector side

$$\dot{S}_D = J_h \frac{U}{T_D} - J_i \quad (11)$$

and the one in the system side

$$\dot{S}_S = -J_e \frac{\Delta\mu}{T_S} - J_h \frac{U}{T_S} + J_i, \quad (12)$$

where $J_i$ is the "information flow",[23] i.e., the transfer rate of entropy from the detector to system dot

$$J_i = J_h \ln \frac{p(x=1|y=0)}{p(x=0|y=0)} \frac{p(x=0|y=1)}{p(x=1|y=1)}. \quad (13)$$

Now, we would like to carefully observe the operating mechanism of the engine, potentially providing new insight into the field. Let $\dot{S}_{DR}$ and $\dot{S}_{SR}$ be the entropy production rates in the reservoirs of the detector and system sides, respectively. When the setup is operated as an information engine, what is happening in the detector side is:

$$\dot{S}_{DR} = \dot{S}_D + J_i, \quad (14)$$

represented by an entropy transfer from the left- to the right-hand side. In the system side, the entropy flow can be described as:

$$J_i = J_e \frac{\Delta \mu}{T_S} + J_h \frac{U}{T_S} + \dot{S}_S \quad (15)$$

and then

$$\dot{S}_S - J_i = \dot{S}_{SR} \quad (16)$$

from the left- to the right-hand side in each equation. (n.b., there is no entropy change in the quantum dots in steady state.) This way, we can see that the entropy increase $\dot{S}_{DR}$ in the detector-side reservoir acts as a source to generate an electrical work $J_e \Delta \mu$ through the system dot. (n.b., Je is defined in the direction against the potential slope $\Delta \mu$.) The electrical work $J_e \Delta \mu$ shares the resource with the heat transfer $J_h U$ and the entropy increase $\dot{S}_S$ in the system side. Thus, we herein define the efficiency of the information engine from Eqs. 14 and 15 as

$$\eta_{IE} \equiv \frac{J_e \Delta \mu}{T_S \dot{S}_{DR}}, \quad (17)$$

which is namely an environment-to-electricity energy conversion efficiency.

3. **Results and Discussion**

Our test calculations exactly reproduced all data series in Figs. 4 and 5 of Ref. 23 thus, verifying the correctness of our numerical calculations. Figure 4 of the present paper shows a set of our calculation results for the output electrical power or work production rate of the information engine comparing the cases $\mu_D = \varepsilon_X - U/2$ (adopted from Ref. 23) and $\mu_D = \varepsilon_X + U/2$. The work production rate is found to be significantly higher for the case $\mu_D = \varepsilon_X + U/2$. This result can be attributed to the faster circulation of $J_h$ and $J_e$ or namely of the states due to the antisymmetric preference of the electronic states (i.e., the states (0, 1) and (1, 0) are designated by the detector dot because of its higher energy level than its reservoir's. See Fig. 3 to understand the states' circulation.). Also important, the condition $\mu_D = \varepsilon_X + U/2$ gives a much wider range of the quantum-dot capacitive interaction strength or the Coulomb interaction energy $U$ for high work production rates and thus would provide a significantly larger tolerance in the setting of $U$ values for high-performance device design and preparation. In contrast, we incidentally observed little differences of the efficiencies $\eta_{IE}$. We thus adopt the condition $\mu_D = \varepsilon_X + U/2$ in the following calculations for the advantage in practical applications of the engines.

Here, we investigate the influence of the energy levels' relative positions among the system dot and its reservoirs. Figure 5 shows the work production rate and efficiency of the information engine with varied $\mu_H$ and $\mu_L$ to $\varepsilon_Y$, keeping $\Delta\mu$ constant. The information-engine efficiency depends little on the relative positions of the energy levels of $\mu_H$ and $\mu_L$ to $\varepsilon_Y$ while the electrical output power or work production rate is somewhat influenced by the relative positions of $\mu_H$ and $\mu_L$ to $\varepsilon_Y$. Sets of $\mu_H$ and $\mu_L$ close to $\varepsilon_Y$ provide higher work production rates, presumably because the large potential slope either between $\mu_H$ and $\varepsilon_Y$ or $\mu_L$ and $\varepsilon_Y$ limits the steady-state current flow through the system dot from the lower to higher reservoir for the cases where $\mu_H$ and $\mu_L$ are far from $\varepsilon_Y$.

Figure 6 shows a set of current–voltage characteristics of the information engine. In these calculations, we always set the position of the energy level of the system dot in the middle of those of it reservoirs, $\varepsilon_Y = (\mu_H + \mu_L)/2$, reflecting the previous result. The difference of the energy levels of the higher- and lower-potential reservoirs of the system dot corresponds to the bias voltage in this electronic setup. From such current–voltage curves, we can graphically recognize the maximum-power point for a set of voltage and current by maximizing the area of the rectangle comprising the point (voltage, current) and the axes as indicated by the shadowed region in Fig. 6, just like for photovoltaic devices.[29] We thereby see that there is a certain value $\Delta\mu$ ($=\mu_H - \mu_L$) which maximizes the electrical output power or work production rate for each condition such as for $U$ determined by the balance of the electrons' flow rate and the potential slope $\Delta\mu$.

We plot the information engine efficiencies $\eta_{IE}$ for various tunneling rates in Fig. 7. We varied the tunneling rates for the electron-flow direction against the potential differences $\Gamma_0^H$ and $\Gamma_1^L$ while fixing the tunneling rates in the direction down the potential slopes $\Gamma_1^H$ and $\Gamma_0^L$. As seen in this set of efficiency results, the efficiency is highly dependent on the antisymmetric relative amplitudes of the electronic tunneling rates, i.e., that $\Gamma_0^H$ and $\Gamma_1^L$ are larger than $\Gamma_1^H$ and $\Gamma_0^L$. It should be noted that we used quite idealistic parameter values in this study for the quantitative investigations: however, for future explorations regarding the realization of real engine devices realistic materials properties for electron transport etc.[44-46] will be needed in calculations and device designing.

Figure 8 shows the efficiency of the information engine $\eta_{IE}$ in dependence on the capacitive interaction strength or quantum-dot Coulomb repulsion energy $U$ for varied temperatures of the detector side $T_D$. The efficiency is found to be strongly dependent on $T_D$ or the relative difference between $T_S$ and $T_D$ and there is a specific $U$ region that yields high $\eta_{IE}$ for each $T_D$. The reason for the existence of optimal $U$ regimes can be attributed to the deficiencies of the electron-transport selectivity for smaller $U$ and the states' circulation velocity for larger $U$. In addition, for the conditions where both $T_D$ and $U$ are small, $\eta_{IE}$ becomes high due to the information domination relative to the heat transfer;[21,23] see Eqs. 10 and 15.

To defend the concept of the environmental information engine presented in this work against the potential suspicion that the engine might in a sense be eventually operated by the heat transfer due to the difference between the two subsystems' temperatures $T_D$ and $T_S$, we compare the information-engine and thermoelectric efficiencies of the setup. We define a thermoelectric efficiency from Eqs. 14 and 15 as:

$$\eta_{TE} \equiv \frac{J_e \Delta\mu}{J_h U}. \quad (18)$$

We then plot in Fig. 9 $\eta_{TE}$ in dependence on $U$ for varied $T_D$'s. Actually, we note from Eqs. 17 and 18 that

$$\frac{\eta_{TE}}{\eta_{IE}} = \frac{T_S}{T_D}. \quad (19)$$

Remarkably, $\eta_{TE}$ thus well exceeds unity for some conditions, verifying that the engine discussed in this study can indeed operate as an environment information engine dominantly driven by the entropy increase in the detector's reservoir $\dot{S}_{DR}$. In other words, the information flow $J_i$ stemming from $\dot{S}_{DR}$ drastically assists the thermoelectric device, resulting in a significant enhancement of $\eta_{TE}$. The operation principle of the engine introduced in this study thus provides new insight towards the utilization of the environmental order as a new energy resource to realize nanodevices that convert the environmental entropy into electrical power.

## 4. Conclusions

In this study, we conducted numerical simulations for a relatively simple model of a type of autonomous coupled-double-quantum-dot information engine. Through the set of calculations, we obtained various design principles for the device parameters, which will be valuable for future device preparation. We introduced a way to see the operation dynamics of the engine, delivering a possibility of a new energy resource of the environmental order, to be converted into electrical power through the information flow.

**Acknowledgment**

We thank Jordan M. Horowitz of Massachusetts Institute of Technology for discussions. This work was partially supported by the Japan Society for the Promotion of Science (JSPS) and the Ministry of Education, Culture, Sports, Science and Technology-Japan (MEXT).

*E-mail: tanabe@cheme.kyoto-u.ac.jp

1) L. Szilard, Z. Phys. **53**, 840 (1929).
2) R. Landauer, IBM J. Res. Dev. **5**, 183 (1961).
3) T. Schreiber, Phys. Rev. Lett. **85**, 461 (2000).
4) R. Kawai, J. M. R. Parrondo, and C. Van den Broeck, Phys. Rev. Lett. **98**, 080602 (2007).
5) O. J. E. Maroney, Phys. Rev. E **79**, 031105 (2009).
6) T. Sagawa and M. Ueda, Phys. Rev. Lett. **102**, 250602 (2009).
7) Y. Fujitani and H. Suzuki, J. Phys. Soc. Jpn. **79**, 104003 (2010).
8) D. Mandal and C. Jarzynski, Proc. Nat'l. Acad. Sci. USA **109**, 11641 (2012).
9) K. Sekimoto and S. Sasa, J. Phys. Soc. Jpn. **66**, 3326 (1997).
10) G. E. Crooks, Phys. Rev. E **60**, 2721 (1999).
11) U. Seifert, Phys. Rev. Lett. **95**, 040602 (2005).
12) J. M. Horowitz and S. Vaikuntanathan, Phys. Rev. E **82**, 061120 (2010).
13) K. Kawaguchi and Y. Nakayama, Phys. Rev. E **88**, 022147 (2013).
14) A. C. Barato and U. Seifert, Phys. Rev. Lett. **112**, 090601 (2014).
15) S. Toyabe, T. Sagawa, M. Ueda, E. Muneyuki, and M. Sano, Nature Phys. **6**, 988 (2010).
16) A. Berut, A. Arakelyan, A. Petrosyan, S. Ciliberto, R. Dillenschneider, and E. Lutz, Nature **483**, 187 (2012).
17) G. Lan, P. Sartori, S. Neumann, V. Sourjik, and Y. Tu, Nature Phys. **8**, 422 (2012).
18) S. Ito and T. Sagawa, Nature Commun. **6**, 7498 (2015).
19) G. Schaller, G. Kieblich, and T. Brandes, Phys. Rev. B **82**, 041303 (2010).
20) M. Esposito and G. Schaller, EPL **99**, 30003 (2012).
21) P. Strasberg, G. Schaller, T. Brandes, and M. Esposito, Phys. Rev. Lett. **110**, 040601 (2013).
22) A. C. Barato, D. Hartich, and U. Seifert, J. Stat. Phys. **153**, 460 (2013).
23) J. M. Horowitz and M. Esposito, Phys. Rev. X **4**, 031015 (2014).
24) D. Hartich, A. C. Barato, and U. Seifert, J. Stat. Mech., P02016 (2014).
25) A. Kutvonen, T. Sagawa, and T. Ala-Nissila, arXiv, 1510.00190 (2015).
26) A. Kutvonen, J. Koski, and T. Ala-Nissila, Sci. Rep. **6**, 21126 (2016).
27) Y. Arakawa and H. Sakaki, Appl. Phys. Lett. **40**, 939 (1982).
28) K. Tanabe, M. Nomura, D. Guimard, S. Iwamoto, and Y. Arakawa, Opt. Express **17**, 7036 (2009).

29) K. Tanabe, D. Guimard, D. Bordel, and Y. Arakawa, Appl. Phys. Lett. **100**, 193905 (2012).
30) K. Ono, D. G. Austing, Y. Tokura, and S. Tarucha, Science **297**, 1313 (2002).
31) T. Fujisawa, T. Hayashi, R. Tomita, and Y. Hirayama, Science **312**, 1634 (2006).
32) G. Schaller, C. Emary, G. Kiesslich, and T. Brandes, Phys. Rev. B **84**, 085418 (2011).
33) J. V. Koski, T. Sagawa, O-P. Saira, Y. Yoon, A. Kutvonen, P. Solinas, M. Moettoenen, T. Ala-Nissila, and J. P. Pekola, Nature Phys. **9**, 644 (2013).
34) J. V. Koski, V. F. Maisi, J. P. Pekola, and D. V. Averin, Proc. Nat'l. Acad. Sci. USA **111**, 13786 (2014).
35) J. V. Koski, V. F. Maisi, T. Sagawa, and J. P. Pekola, Phys. Rev. Lett. **113**, 030601 (2014).
36) J. V. Koski, A. Kutvonen, I. M. Khaymovich, T. Ala-Nissila, and J. P. Pekola, Phys. Rev. Lett. **115**, 260602 (2015).
37) C. Van den Broeck, Phys. Rev. Lett. **95**, 190602 (2005).
38) T. Sagawa and M. Ueda, Phys. Rev. Lett. **100**, 080403 (2008).
39) M. Esposito, K. Lindenberg, and C. Van den Broeck, Phys. Rev. Lett. **102**, 130602 (2009).
40) K. Kawaguchi and M. Sano, J. Phys. Soc. Jpn. **80**, 083003 (2011).
41) R. Sanchez and M. Buttiker, Phys. Rev. B **83**, 085428 (2011).
42) J. M. Horowitz and K. Jacobs, Phys. Rev. E **89**, 042134 (2014).
43) N. Shiraishi, Phys. Rev. E **92**, 050101 (2015).
44) C. W. J. Beenakker, Phys. Rev. B **44**, 1646 (1991).
45) W. G. van der Wiel, S. De Franceschi, J. M. Elzerman, T. Fujisawa, S. Tarucha, and L. P. Kouwenhoven, Rev. Mod. Phys. **75**, 1 (2003).
46) W. Lu, Z. Ji, L. Pfeiffer, K. W. West, and A. J. Rimberg, Nature **423**, 422 (2003).

**Figure Captions**

Fig. 1. (Color online) Schematic illustration of the model autonomous coupled-double-quantum-dot information engine.

Fig. 2. (Color online) Schematic diagram of the relationship among the potential energy levels in the components of the information-engine model setup.

Fig. 3. (Color online) Schematic diagram of the electronic states $(x, y)$. $J_e$ and $J_h$ are the electric current and heat flow, respectively.

Fig. 4. (Color online) Electrical output power $J_e\Delta\mu$ representing the work production rate of the information engine in dependence on the capacitive interaction strength between the quantum dots $U$ for various temperatures of the detector quantum dot $T_D$ for the cases (a) $\mu_D = \varepsilon_X - U/2$ and (b) $\mu_D = \varepsilon_X + U/2$ under the condition $\varepsilon_X = \varepsilon_Y = 1$, $\mu_H = 1.1$, $\mu_L = 0.9$, $T_S = 1$, $\Gamma = 100$, $\Gamma_0^H = \Gamma_1^L = 10$, and $\Gamma_1^H = \Gamma_0^L = 0.1$.

Fig. 5 (Color online) (a) Electrical output power $J_e\Delta\mu$ representing the work production rate and (b) efficiency $\eta_{IE}$ of the information engine in dependence on the capacitive interaction strength between the quantum dots $U$ for various relative energy-level positions $\mu_H$ and $\mu_L$ to $\varepsilon_Y$, keeping $\Delta\mu = 0.2$ constant, under the condition $\varepsilon_X = \varepsilon_Y = 1$, $T_D = 0.1$, $T_S = 1$, $\Gamma = 100$, $\Gamma_0^H = \Gamma_1^L = 10$, and $\Gamma_1^H = \Gamma_0^L = 0.1$.

Fig. 6. (Color online) Current–voltage characteristics of the information engine with various capacitive interaction strengths between the quantum dots $U$, keeping $\varepsilon_Y = (\mu_H + \mu_L)/2$, under the condition $\varepsilon_X = \varepsilon_Y = 1$, $T_D = 0.1$, $T_S = 1$, $\Gamma = 100$, $\Gamma_0^H = \Gamma_1^L = 10$, and $\Gamma_1^H = \Gamma_0^L = 0.1$. The shadowed rectangular area represents the maximum output power $P_{MAX}$ of the engine for $U = 0.6$.

Fig. 7. (Color online) Information-engine efficiency $\eta_{IE}$ in dependence on the capacitive interaction strength between the quantum dots $U$ for various tunneling rates $\Gamma_0^H$ and $\Gamma_1^L$, under the condition $\varepsilon_X = \varepsilon_Y = 1$, $\mu_H = 1.1$, $\mu_L = 0.9$, $T_D = 0.1$, $T_S = 1$, $\Gamma = 100$, and $\Gamma_1^H = \Gamma_0^L = 0.1$.

Fig. 8. (Color online) Information-engine efficiency $\eta_{IE}$ in dependence on the capacitive interaction strength between the quantum dots $U$ for various temperatures of the detector quantum dot $T_D$, under the condition $\varepsilon_X = \varepsilon_Y = 1$, $\mu_H = 1.1$, $\mu_L = 0.9$, $T_S = 1$, $\Gamma = 100$, $\Gamma_0^H = \Gamma_1^L = 10$, and $\Gamma_1^H = \Gamma_0^L = 0.1$.

Fig. 9. (Color online) Thermoelectric efficiency $\eta_{TE}$ in dependence on the capacitive interaction strength between the quantum dots $U$ for various temperatures of the detector quantum dot $T_D$, under the condition $\varepsilon_X = \varepsilon_Y = 1$, $\mu_H = 1.1$, $\mu_L = 0.9$, $T_S = 1$, $\Gamma = 100$, $\Gamma_0^H = \Gamma_1^L = 10$, and $\Gamma_1^H = \Gamma_0^L = 0.1$.

**Figures**

Figure 1

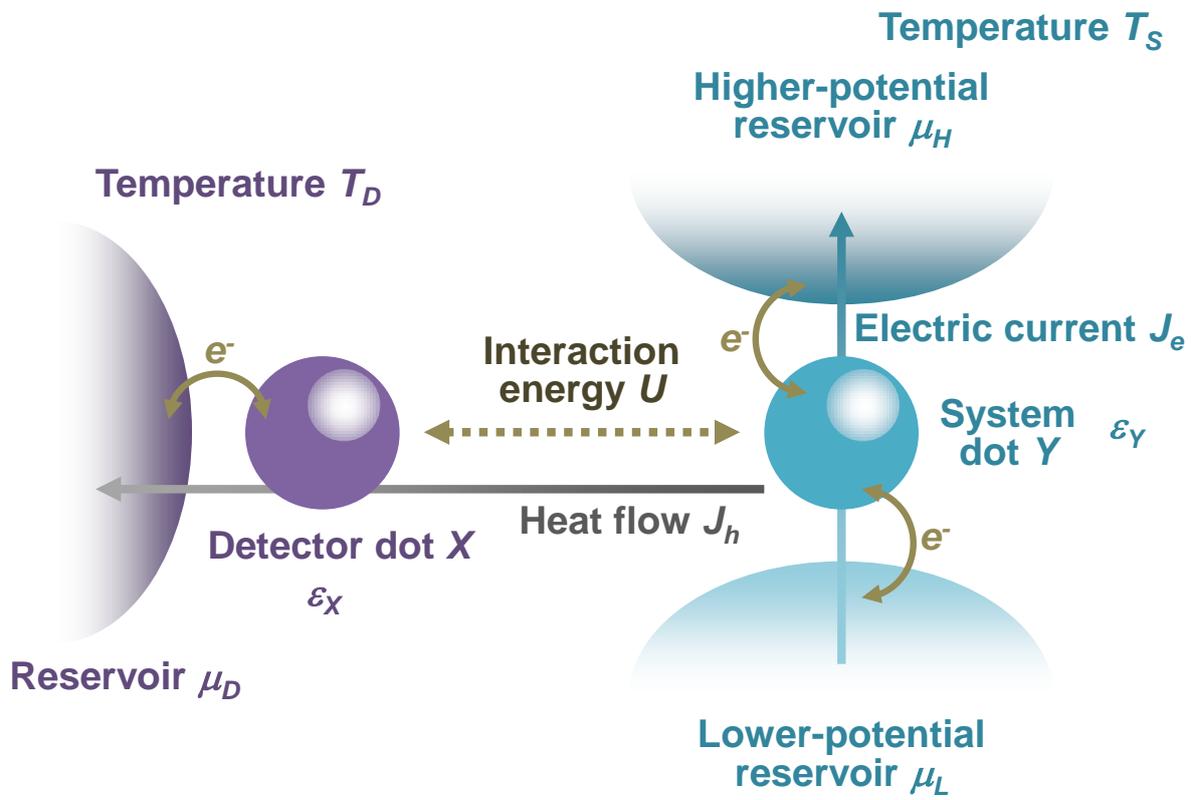

Figure 2

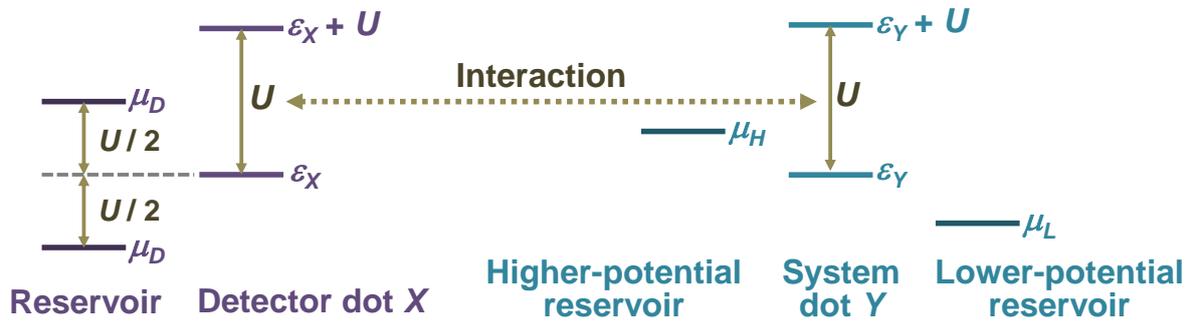

Figure 3

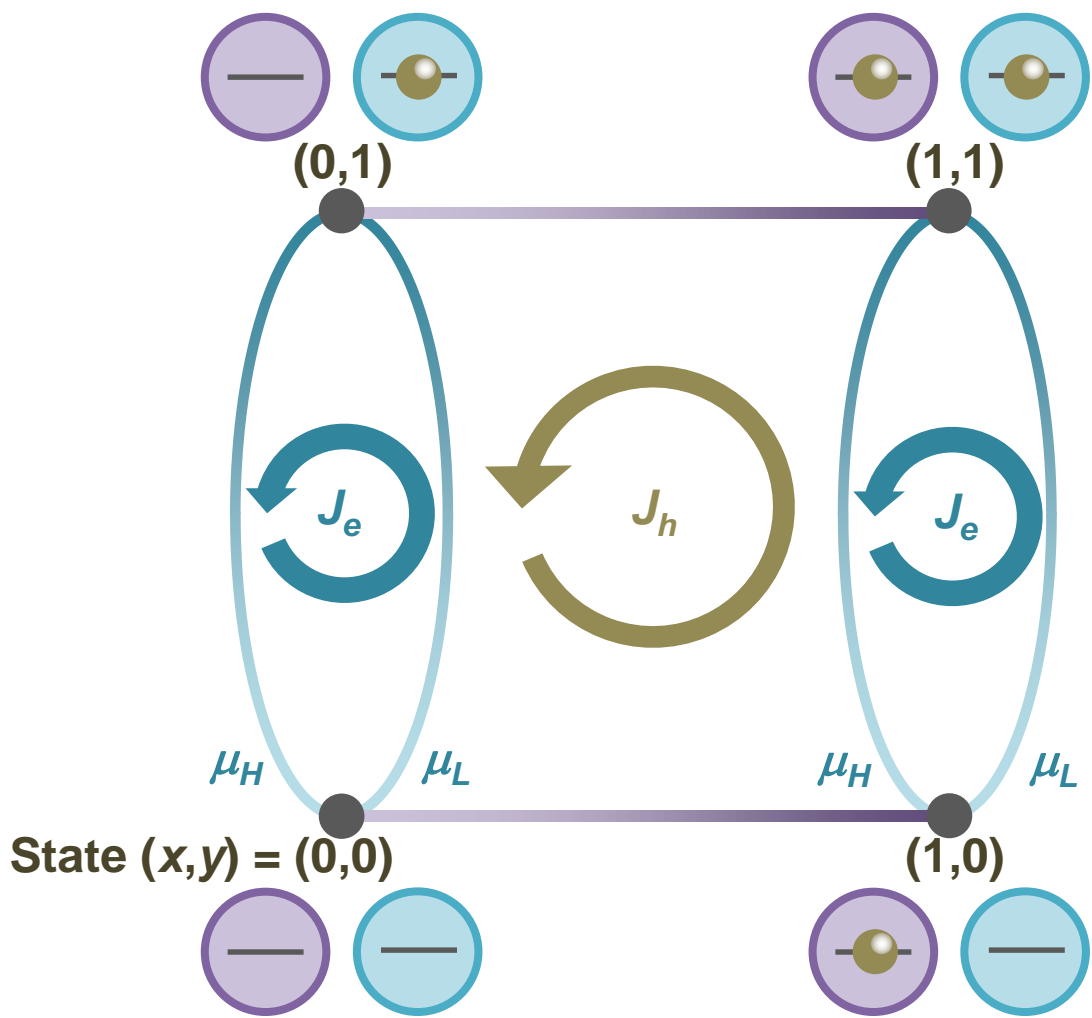

Figure 4

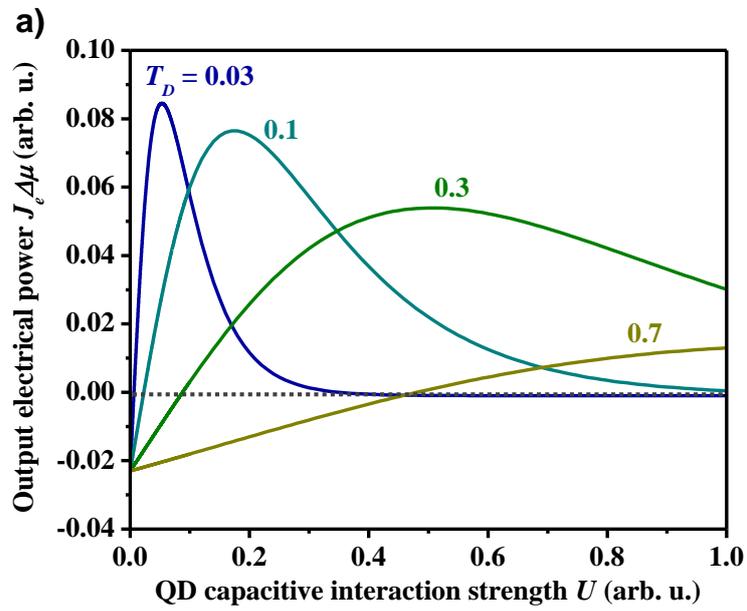

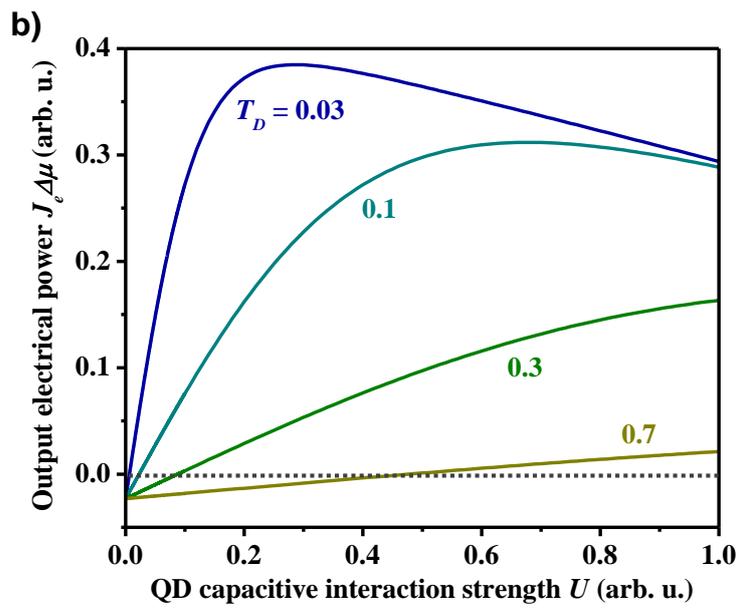

Figure 5

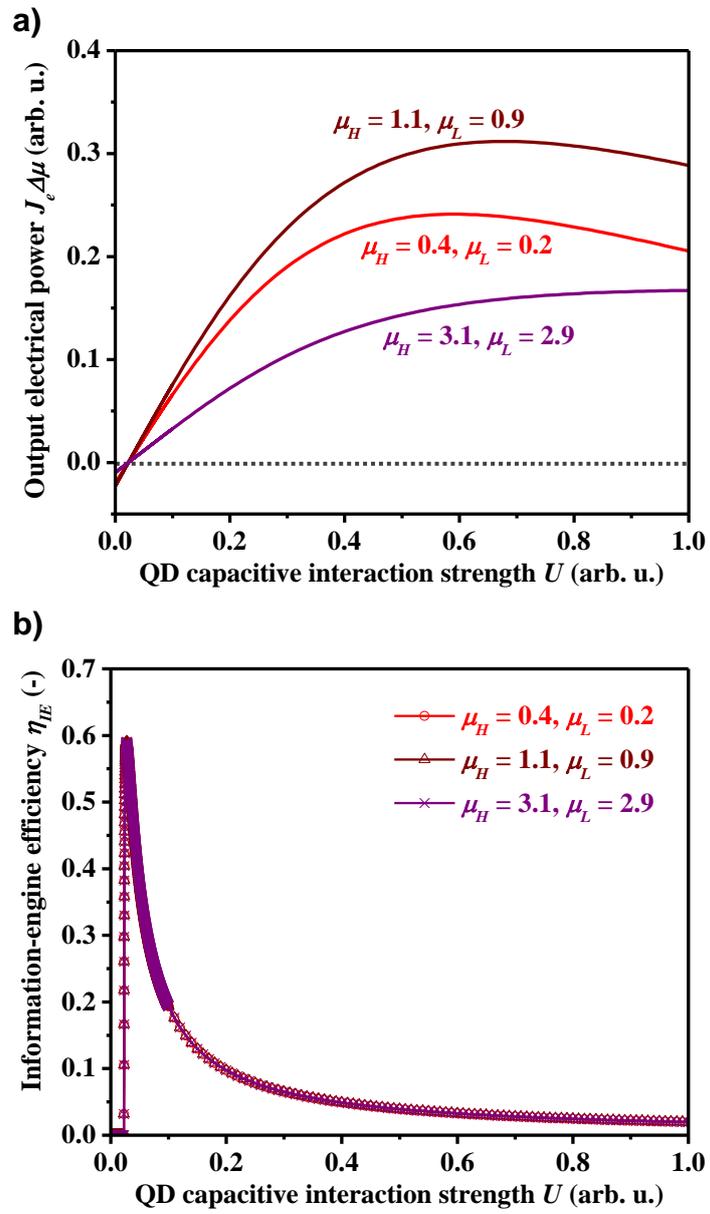

Figure 6

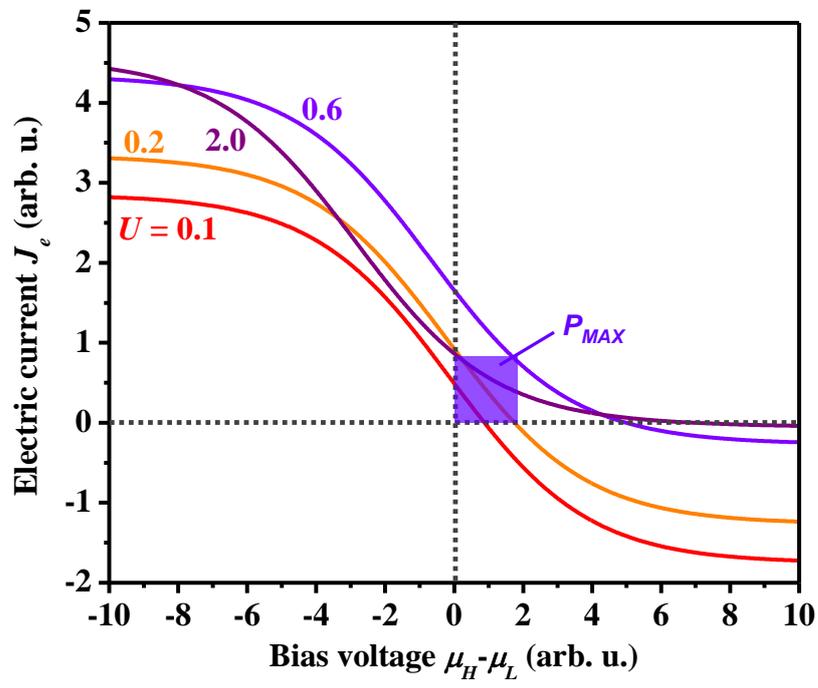

Figure 7

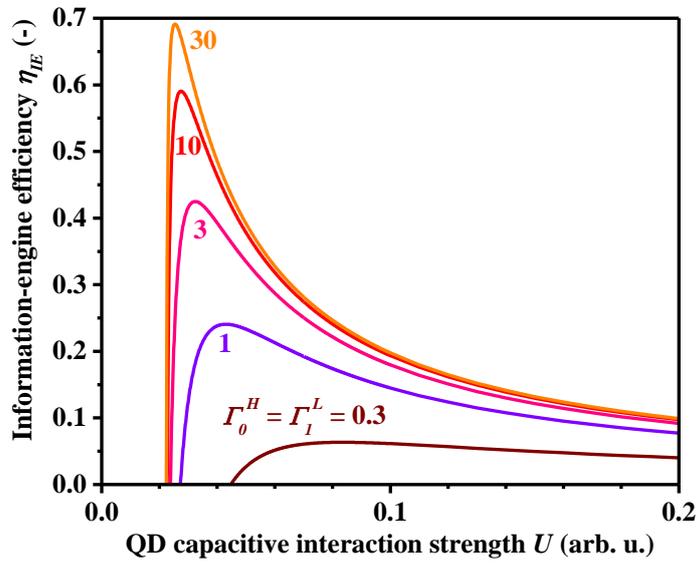

Figure 8

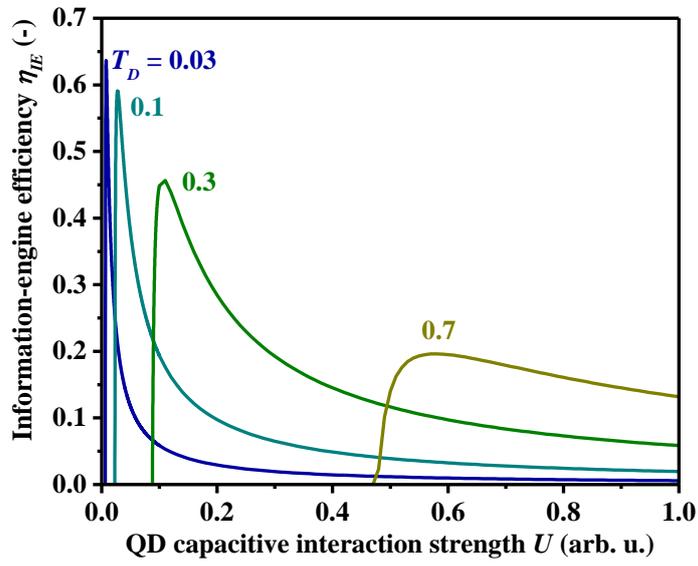

Figure 9

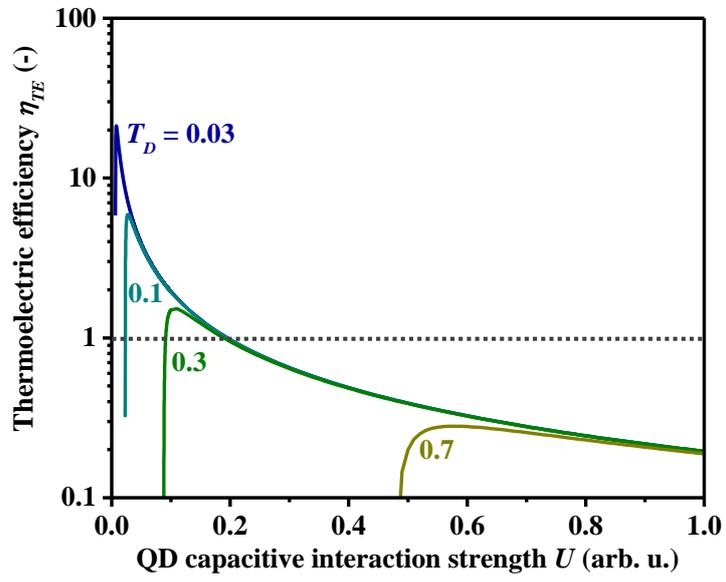